\documentstyle[12pt,aaspp]{article}

\begin{document}
\title{AMPLITUDE OF PRIMEVAL FLUCTUATIONS FROM COSMOLOGICAL MASS DENSITY
RECONSTRUCTIONS}
\author{Uro\v s Seljak\altaffilmark 1 and Edmund Bertschinger}
\affil{Department of Physics, MIT, Cambridge, MA 02139 USA}
\begin{quote}
\altaffilmark{}
\altaffiltext{1}{Also Department of Physics, University of Ljubljana,
Slovenia}
\end{quote}

\begin{abstract}
We use the POTENT reconstruction of the mass density field in the
nearby universe to estimate the amplitude of the density
fluctuation power spectrum for various cosmological models.
We find $\sigma_8\,\Omega_m^{0.6}=1.3^{+0.4}_{-0.3}$, almost
independently of the power spectrum.
This value
agrees well with the {\sl COBE} normalization for the standard CDM
model, while some alternative
models predict an excessive amplitude compared with {\sl COBE}.
Flat low $\Omega_m$ models and tilted models with spectral index
$n<0.8$ are particularly discordant.
\end{abstract}


\def\ltsima{$\; \buildrel < \over \sim \;$}
\def\lsim{\lower.5ex\hbox{\ltsima}}
\def\gtsima{$\; \buildrel > \over \sim \;$}
\def\gsim{\lower.5ex\hbox{\gtsima}}

\section{Introduction}
A central problem in cosmology today is the nature and abundance
of the matter in the universe.  While a given matter content
uniquely determines the shape of the power spectrum
of density fluctuations (up to variations in the primeval shape),
the amplitude of mass fluctuations remains unspecified theoretically and its
value must be sought from observations.
After the {\sl COBE} discovery of primeval fluctuations
(\cite{smoot92}) it became possible
to use temperature fluctuations at the surface of last
scattering to
compute the small scale normalization $\sigma_8$ (the rms relative mass
fluctuation in a sphere of radius $8\,h^{-1}\,{\rm Mpc}$,
$H_0=100$ h km s$^{-1}$ Mpc$^{-1}$) for any particular
model (\cite{wright92}; \cite{efstathiou92}; \cite{adams93}).
This allowed researchers to shift
their emphasis from the amplitude determination to the study of the spectral
shape.  Any additional independent amplitude determinations would
enable one to discriminate between different models of structure formation.
The {\sl COBE} results are able to provide a constraint on the shape of the
primordial power spectrum, but not on the matter content of the universe,
because on the scales probed by {\sl COBE} different matter contents
produce similar temperature fluctuations. Small scale CMB anisotropy
experiments
are just beginning to be useful for determining the spectral shape
(e.g., Bond et al. 1991;
Dodelson \& Jubas 1993; G\'orski, Stompor \& Juszkiewicz 1993).

An alternative approach is to use measured redshift-distance samples
to determine the amplitude of mass fluctuations. This
method has the advantage that the reconstructed peculiar velocities
are directly sensitive to the underlying mass distribution.
Most work to date has been based on estimates of the  bulk flow averaged
over large volumes (Bertschinger et al. 1990; Courteau et al. 1993).
These statistics have several limitations.  First, the statistical
distribution of spectral amplitude estimates based on the bulk flow
is broad ($\chi^2$ with only 3 degrees of freedom) and consequently
bulk flows can only weakly constrain the models.  Second, bulk flows
are particularly sensitive to the systematic errors introduced by
nonuniform sampling (the sampling gradient bias of Dekel, Bertschinger,
\& Faber 1990, hereafter DBF).  Finally, the scales contributing to the
bulk flow estimates are large ($\gsim40$--60 $h^{-1}$ Mpc), whereas
most of the peculiar velocity measurements come from smaller distances
(10--30 $h^{-1}$ Mpc).  It should be possible to place additional
independent model constraints on the smaller scales alone.
Several groups have combined different peculiar velocity data to
probe models on a range of scales (e.g., Del Grande \& Vittorio 1992;
Muciaccia et al. 1993; Tormen et al. 1993), but all except
Tormen et al. use bulk flow estimates with their associated uncertainties.

In this paper we estimate the amplitude of mass fluctuations on
intermediate scales by applying the maximum-likelihood method directly to the
POTENT
reconstruction of density perturbations from the peculiar velocities
(Bertschinger \& Dekel 1989; DBF).
This analysis assumes that the measured velocities are a fair tracer
of the underlying velocity field, which was induced by the underlying
gravitational field.
Use of a large sample of peculiar velocity data reduces
both the statistical and systematic errors, provided that care is
exercised concerning nonlinear corrections and Malmquist bias effects.
We test several models: standard cold dark matter (CDM), CDM plus a
cosmological constant(CDM+$\Lambda$), and CDM plus massive neutrinos
(CDM+HDM), with several different choices for the Hubble constant and
the primeval spectral index.
Because we use the relatively sparse 1990 dataset we do not
attempt to discriminate between the models using the velocity data alone.
Instead,
we compare our results to the {\sl COBE} normalization to derive some
conclusions about the viability of the models. We also briefly
comment on the
agreement with other methods of amplitude estimation.

\section{Method and data analysis}
Assuming a potential flow the present velocity field can be extracted from
observed radial peculiar velocities of galaxies (e.g. Lynden-Bell et al.
1988) by integrating along radial rays. This is the essence of the
POTENT reconstruction method (Bertschinger \& Dekel 1989; DBF).
Furthermore, in linear perturbation theory there is a simple relation
between velocity and density fields (Peebles 1980): $\delta({\bf
r})=-(H_0f)^{-1}\nabla\cdot {\bf v(r)}\equiv
f^{-1}\widetilde{\delta}({\bf
r})$, where $\delta({\bf r})$ is
the density fluctuation at position ${\bf r}$, $H_0$ is the Hubble
constant and $f$ is the the growing mode logarithmic growth rate.
For CDM and CDM+$\Lambda$ models at low redshifts f is well approximated
by $f(\Omega_m)=\Omega_m^{0.6}$ (Peebles 1980;
Lahav et al. 1991), where $\Omega_m$ is the total matter content in the
universe. For the CDM+HDM model $f$ is in general a function of the
wavenumber $k$
(see section 3.2). We introduce $\widetilde{\delta}({\bf r})$
as the measured quantity to be compared with the theoretical predictions.

The input data and their treatment are described in Bertschinger et al.
(1990).  We discard all points with $r > 50h^{-1}$  Mpc
and with the distance to the fourth nearest neighbor $R_4>10h^{-1}$ Mpc,
thus keeping only the points with small sampling and measurement errors.
At the end we are left with $N=111$ data points $\widetilde{\delta_i}$
on a grid of spacing $10h^{-1}$  Mpc (with gaussian smoothing
radius $12h^{-1}$  Mpc),
which we use for comparison with the theoretical predictions.
Despite the many
data points there are only about 10 independent degrees of freedom in the
sample (Dekel et al. 1993), which currently prevents us from extending
the analysis beyond the amplitude determination.

Given the data one can estimate the power spectrum parameters using the
maximum likelihood method. The initial density perturbations are assumed to
constitute a gaussian random field in all the models that we study here.
Nonlinear effects and nontrivial coupling between signal and
noise (sampling and measurement errors)
in the data affect the distributions and in general the resulting data are
not normally distributed. However, one can still define a gaussian
likelihood function and use its maximum as a statistic
to estimate the unknown parameters. Moreover, if the deviations from
a normal distribution are small, the increase in variance owing to the use
of the wrong likelihood function should be small.  We use Monte Carlo
simulations with the correct distributions to estimate the bias and variance
of the estimated parameters.

We define the likelihood function as
\begin{equation}
L(\sigma_8)={1 \over \sqrt{(2\pi)^N \vert M\vert}}
\exp\left[ -{1\over 2}\sum_{i=1}^N \sum_{j=1}^N
M_{ij}^{-1}(\widetilde{\delta_i}-\langle \widetilde{\delta_i} \rangle)(
\widetilde{
\delta_j}-\langle \widetilde{\delta_j}
\rangle) \right],
\label{likest}
\end{equation}
where $M_{ij}^{-1}$ and $M$ are the inverse and determinant of
the correlation matrix $M_{ij}=\langle (\widetilde{\delta_i}-\langle
\widetilde{
\delta_i}
\rangle)(\widetilde{\delta_j}-\langle \widetilde{\delta_j} \rangle) \rangle$.
Here $\langle \rangle$ denotes averaging over the random field
ensemble (signal) and distance errors (noise).
For a given theoretical model, the correlation matrix depends on the
parameters of the power spectrum, most notably on its amplitude. An
estimate of the amplitude $\sigma_8$ is given by the value that maximizes
$L$.  In general this estimate is biased, in part because the maximum
likelihood estimator is only asymptotically unbiased,
but also because of our assumption of normality and owing to the presence
of sampling gradient and other biases in the $\{\widetilde{\delta_i}\}$.
We estimate the statistical bias by Monte Carlo simulations
and correct for it as described below.

To apply the method in practice, one needs to compute the correlation matrix
$M_{ij}$. The matrix has contributions from both the noise and the true
underlying signal. The signal contribution is given by
\begin{equation}
M_{ij}^s=\int d^3k\,f^2(\Omega_m,k)\,W^2(k)\,P(k)\,e^{i{\bf k}\cdot
({\bf r}_i-{\bf r}_j)}\ ,
\end{equation}
where $P(k)$ is the density fluctuation power spectrum and $W(k)$
is the smoothing window function in $k$-space.
Note that we have retained a possible $k$ dependence of $f$.

Measurement noise arises from errors in the galaxy
distance estimates and from the sparse spatial sampling of the data. A detailed
analysis, performed in the appendix of DBF, shows that the noise contribution
is correlated with the signal contribution in $M_{ij}$. Thus,
one cannot compute each of the two contributions separately
and then add them together with the appropriate amplitude, rather, one must
compute $M_{ij}$ directly for each amplitude.
We computed $M_{ij}$ using Monte Carlo simulations as described by
DBF and Bertschinger et al. (1990), with input peculiar velocities given
by a random sample of the linear velocity field from a particular gaussian
theory (e.g., CDM).  We evaluated the radial velocities at the positions
of real galaxies and added noise assuming a lognormal distribution of
distance errors (in contrast with the normal distribution used in the
earlier work).  For each model and for a dense grid of amplitudes,
500 POTENT reconstructions were averaged over the signal and noise
ensembles to compute $\langle \widetilde\delta_i\rangle$ and $M_{ij}$.

Before discussing the results we have to address the possible effects
of bias $\langle \widetilde\delta_i \rangle$. As discussed in DBF,
the two main contributions are from
the sampling gradient bias and the Malmquist bias.
The first one is adequately taken into account by the Monte Carlo
simulations,
because they use similar sampling of space as the real data.
On the other hand, Malmquist bias computed by the Monte Carlo simulations
is not exactly the real bias present in the data.
Since the real data had the homogeneous Malmquist bias subtracted,
the only bias left
is the density gradient bias $(\partial \ln n /\partial \ln r)
(\sigma^2/ r) $. However, the bias in the Monte Carlo samples is
$(3.5+\partial \ln nP /\partial \ln r)
(\sigma^2/ r) $, where $P(\vec r\,)$ is a poorly known selection function.
Without knowledge of $P$, one cannot properly correct the simulations
for the Malmquist
bias. To test the importance of inhomogeneous Malmquist bias on the
parameters we wish to
estimate, we compare the results for two simple cases. In first case, we
correct the real data
for the Monte Carlo bias, while in the second we do not.
This is a correct treatment of the bias if the
density gradient bias is negligible in the first case or the
volume and selection function contributions to the bias
exactly cancel in the second case.
We find that the final mass amplitudes
differ from each
other by about 10\%.  This is significantly smaller than the statistical
errors of our amplitude estimates.

Another possible source of error are nonlinear effects, which tend to
change the densities and velocities compared to their linear values.
The real velocities are nonlinear while our simulations are strictly
linear.
Due to the large smoothing radius, the differences are small in the
case studied here, even if the unsmoothed density fluctuations are large.
To test this one can also compute the true density field
by applying the following correction
in the quasi-linear regime ($-0.8\lsim\delta\lsim4.5$):
$f^{-1}\widetilde{\delta}=\delta/(1+0.18\delta)$
(Nusser et al. 1991).
This changes our final estimated amplitudes by 5--10\%, depending on the value
of $f$.
Note that inhomogeneous Malmquist bias tends to increase the amplitude
of the measured peculiar velocities, while nonlinear effects decrease the
linear $\widetilde\delta$.  Because both effects are small ($\sim10\%$)
and opposite in sign, we will neglect them.

\section{Results}
Our analysis was restricted to the simplest
generalizations of standard CDM model that decrease the power on
small scales relative to that on large scales and have been proposed recently
as viable models of large scale structure.
We computed transfer functions for the models by integrating the coupled
linearized relativistic Einstein, Boltzmann, and fluid equations for
baryons, CDM, photons, and neutrinos.
In all cases, we fix the baryon contribution to
$\Omega_B=0.0125h^{-2}$, as given by the nucleosynthesis constraint
(Walker et al. 1991).

For each model considered, we estimated the maximum-likelihood value
of the amplitude, which we denote $\sigma_{8,v}$.
The error distribution of estimated $\sigma_{8,v}$ from the Monte Carlo
samples is somewhat asymmetric with longer
tails toward larger values (fig. \ref{fig1}).
For all the models,
the relative one-sigma errors are well approximated by $(+0.3/-0.25)$.
To this one must add the sum of systematic errors due to the Malmquist
bias in the data and to residual nonlinear effects, whose sign is unclear
but whose magnitude is, at most, about 10\%.

The results for different models
are summarized in Table \ref{table1}.  Our
estimates  of $\sigma_{8,v}$ from POTENT include a small statistical bias
correction ($\sim5\%$). In addition, in Table \ref{table1}
we include $\sigma_{8,l=2}$,
the {\sl COBE} normalization based on assuming a value of
$Q_{\rm rms-PS}=15.7\exp[0.46(1-n)]\,\mu$K
(\cite{smoot92}; \cite{seljak93}).
We also include the age of the universe as a possible
additional constraint for these models.
Despite the complicated window function,
we find that there is only a weak dependence of $\sigma_{8,v}$ on the shape
of the power spectrum. This is not surprising considering that the
POTENT sample is sparse at large distances and has been smoothed
with a gaussian of radius $12\,h^{-1}\,{\rm Mpc}$.
We find that $\sigma_{8,v}$ depends mainly on $\Omega_m$ through
$f(\Omega_m)$.  An approximate value valid through most of the parameter
space is thus
\begin{equation}
\sigma_{8,v}\,\Omega_m^{0.6}=1.3^{+0.4}_{-0.3}\ .
\label{eqn1}
\end{equation}
95\% confidence limit intervals give $\sigma_8\,\Omega_m^{0.6}\sim 0.7-2.3$.
For a given model (and thus, for a given value of $n$), the relative
1$\sigma$ uncertainty of $\sigma_{8,l=2}$ is only $17\%$ (\cite{seljak93};
Scaramella \& Vittorio 1993), which is 2 times smaller than the uncertainty
of $\sigma_{8,v}$. As a first approximation
one can thus neglect the uncertainty of $\sigma_{8,l=2}$ when
comparing it with $\sigma_{8,v}$.

\begin{table}[p]
$$
%\begin{array}{cccccccc}
\begin{array}{llllllll}
\Omega_{CDM+B} & \Omega_{\Lambda} & \Omega_{\nu} & h & n & \sigma_{8,v} &
\sigma_{8,l=2} & {\rm age [Gyr]}\\
\hline
1.0 & 0.0 & 0.0 & 0.5 & 1.0 & 1.4 & 1.05 & 13.1\\
0.2 & 0.8 & 0.0 & 0.8 & 1.0 & 3.0 & 0.67 & 13.2\\
0.7 & 0.0 & 0.3 & 0.5 & 1.0 & 1.3 & 0.66 & 13.1\\
0.8 & 0.0 & 0.2 & 0.75 & 1.0 & 1.3 & 1.20 & 8.7\\
1.0 & 0.0 & 0.0 & 0.5 & 0.75& 1.3 & 0.55 & 13.1\\
\hline
\end{array}
$$
\caption{$\sigma_{8,v}$, $\sigma_{8,l=2}$ and age of the universe
for various models discussed
in the text. Relative errors are $^{+0.4}_{-0.3}$ for $\sigma_{8,v}$
and $\pm 0.17$ for $\sigma_{8,l=2}$.}
\label{table1}
\end{table}
\subsection{CDM model}

The first model we tested is
the CDM model. We performed Monte Carlo simulations for the standard
case with $h=0.5$. The $\sigma_{8,v}$ and $\sigma_{8,l=2}$
amplitudes agree well with each other (see Table \ref{table1}). In fact,
the amplitude predicted by the
standard CDM model is consistent with the {\sl COBE} normalization over most
of the allowed range of $h$. Thus, for example, $\Omega_m=1$ and $h=0.75$
(not shown in the Table) gives
$\sigma_{8,l=2}=1.5$, which is still compatible within the uncertainties
with $\sigma_{8,v}=1.3$.
The CDM model cannot be ruled out based on the comparison
between the velocity data and {\sl COBE}. This conclusion
agrees with the previous
comparisons based on the bulk flow estimates on somewhat larger scales
(Bertschinger et al. 1990; Efstathiou et al. 1992; Courteau et al. 1993),
but the present analysis
gives smaller uncertainty in the amplitude and thus places more
stringent limits on the models.

\subsection{CDM+$\Lambda$ models}

We studied a family of generalized CDM models adding
a cosmological constant $\Lambda$ with $\Omega_m+\Omega_{\Lambda}=1$.
Monte Carlo simulations were performed for
the model with $h=0.8$ and
$\Omega_{\Lambda}=0.8$. This model is a representative of the models
which match the recent determinations
of large scale galaxy clustering power (Maddox et al. 1990;
Kofman, Gnedin \& Bahcall 1993)
and the value of $h$ (Jacoby et al. 1992),
yet is compatible with globular cluster ages.
Our values for $\sigma_{8,l=2}$  are lower than the corresponding values
given by Efstathiou et al. (1992) owing to a different baryon content (which
affects the density transfer function)
and because we include the contribution
from the time derivative of the potential integrated along the line of sight
when $\Lambda \ne 0$ (\cite{kofman85}; \cite{gorski92}).
The agreement with the low $\Omega_m$ model is
not very good. While decreasing $\Omega_m$ increases $\sigma_{8,v}$
[due to the $f(\Omega_m)$ factor],
the time derivative of the potential integrated along the line of
sight tends to decrease $\sigma_{8,l=2}$ for a given $\Delta T/T$
quadrupole.
Decreasing $h$
even further decreases $\sigma_{8,l=2}$. We find that the results
strongly constrain these models toward small
$\Omega_{\Lambda}$ values.
95\% upper limits on $\Omega_{\Lambda}$ are 0.6 for $h=0.8$ and 0.4
for $h=0.5$.

\subsection{CDM+HDM models}

A mixed dark matter model with $\Omega_{CDM+B}=0.7$
and $\Omega_{\nu}=0.3$ has recently emerged as one of the best
candidates to explain the large-scale structure measurements
(Schaefer, Shafi, \& Stecker 1989;
Davis, Summers, \& Schlegel 1992; Taylor \& Rowan-Robinson 1992;
Klypin et al. 1993).
In this model, the
growth factor $f(k)$ depends on wavenumber because free-streaming damps
small wavelengths; $f(k)$ ranges between 1 on large scales to
${1\over 4}[(1+24\Omega_{CDM+B})^{1/2}-1]$ on small scales
(Bond, Efstathiou, \& Silk 1980; Ma 1993).
For $h=0.5$, which corresponds to $m_{\nu}=7$ eV, $f=1$ on large scales
and $f\approx 0.8$ on small scales, with a transition at
$k\approx1\,{\rm Mpc}^{-1}$.  Because this is a relatively small scale,
the effect on our estimate of $\sigma_{8,v}$ is small.

The results for this model in Table \ref{table1} imply that the CDM+HDM
model is not strongly constrained,
although the estimated $\sigma_{8,v}$ is
somewhat high compared to the $\sigma_{8,l=2}$.
Decreasing the $\Omega_{\nu}$
contribution or increasing the value of $h$ increases
$\sigma_{8,l=2}$ and thus reduces the discrepancy.
We find excellent agreement between the two normalizations
for a particular model of $\Omega_{\nu}=0.2$ and $h=0.75$, but
of course other parameter values with smaller $\Omega_{\nu}$
and/or larger $h$ give acceptable results as well. In general,
the difference between CDM and CDM+HDM power spectra is small on large
scales ($>8 h^{-1}$Mpc) and consequently the {\sl COBE}
normalized $\sigma_{8,l=2}$ differs little for the two models.

\subsection{Tilted models}

A third way to decrease small scale power relative to that on large scales
is to tilt the
primordial power spectrum $P(k)\propto k^n$ by decreasing $n$
(\cite{adams93}; \cite{muciaccia93}).
The power spectrum for tilted models
differs from its standard CDM counterpart ($n=1$) on all scales,
not just on small scales as for the CDM+HDM power spectra.
Because the {\sl COBE}
scale is about 3 orders of magnitude larger than the $\sigma_8$ scale,
a given $Q_{\rm rms-PS}$ normalization drastically
changes $\sigma_{8,l=2}$ even for modest changes in $n$. This is
true despite the fact that
the best-fit value of $Q_{\rm rms-PS}$ increases when $n$ is decreased
(\cite{seljak93}). For the CDM transfer function with $h=0.5$ we find
\begin{equation}
\sigma_{8,l=2}=1.05\,(1\pm0.17)e^{-2.48(1-n)},
\label{adams}
\end{equation}
similar to the expression based on the $10^{\circ}$ {\sl COBE} normalization
(Adams et al. 1993).

Low values of $n$ generally imply an excessively small value of
$\sigma_{8,l=2}$ relative to $\sigma_{8,v}$.
For the particular case $n=0.75$ and $h=0.5$, the discrepancy between
$\sigma_{8,l=2}$ and $\sigma_{8,v}$ is more than a factor
of 2 (see Table \ref{table1}; the {\sl COBE} normalization assumes
no gravitational wave contribution). Including a possible
gravitational wave contribution (Lucchin, Matarrese, \& Mollerach 1992;
Davis et al. 1992) further decreases $\sigma_{8,l=2}$,
by a factor of $\sqrt{(3-n)/(14-12n)}$.
Higher values of $h$ allow somewhat lower values of $n$,
but in general $n$ cannot differ from 1 by more than $\sim 0.1$-$0.2$.
95\% confidence limits give $n>0.85$ with no gravitational wave contribution
and $n>0.94$ with a gravitational wave contribution, assuming $h=0.5$.

\section{Summary}
The analysis presented here gives the amplitude of mass
fluctuations $\sigma_{8,v}\approx1.3\,\Omega_m^{-0.6}$ in different models.
Comparing the $\sigma_{8,v}$ with the {\sl COBE} normalized
$\sigma_{8,l=2}$, one
can constrain different models of structure formation,
due to the fact that the two normalizations work on very different scales.
In general, all the usual extensions of CDM, which decrease the power on
small scales relative to that on
large scales, decrease the $\sigma_{8,l=2}$ value
relative to the standard CDM value, but leave $\sigma_{8,v}$ almost
unchanged.
While the estimated $\sigma_{8,v}$
agrees well with the {\sl COBE} value for standard CDM, the agreement
becomes worse for the extensions of CDM.
This particularly strongly challenges
the non-zero cosmological constant models and the tilted
models. It also points to a somewhat lower value of
$\Omega_{\nu}$ or a higher value of $h$
than have been assumed by most workers studying the mixed dark matter models.

Recently, using the constraints from the masses and abundances
of rich clusters, Efstathiou, White, \& Frenk (1993) obtained
$\sigma_8=0.57\,\Omega_m^{-0.56}$. This result is inconsistent with our
result at more than the 2$\sigma$ level, although including all the
sources of systematic errors could bring the two results into better
agreement.  Nevertheless, there is increasing evidence that the
estimate of $\sigma_8$ from cluster abundances give lower values
than estimates based on peculiar velocities (Lilje 1992;
Henry \& Arnaud 1991; Evrard 1989).
%; but see Bond \& Myers 1992).
The discrepancy may point to a significant
systematic error in either of the two methods. Further investigations
are needed to resolve this issue.

\acknowledgments
We thank Avishai Dekel for allowing us the use of the POTENT results.
This work was supported by grants NSF AST90-01762 and NASA NAGW-2807.

\newpage
%\listoffigures
\begin{figure}[p]
\vspace*{16 cm}
\caption{Distribution of $\sigma_8$ estimates from the 500 Monte Carlo
simulations of the standard CDM power spectrum. The
input amplitude value is $\sigma_8=1.5$.  The statistical bias correction
to $\sigma_8$ given by the above distribution is $-0.1$.
Other models and amplitudes give similar error distributions.}
\label{fig1}
\end{figure}

\end{document}